# Active multi-point microrheology of cytoskeletal networks

**Tobias Paust[2,3], Katinka Mertens[1], Ines Martin[1], Michael Beil[2], and Othmar Marti[1]**

[1]Institute of Experimental Physics, Ulm University, Albert-Einstein-Allee 11, 89081 Ulm, Germany
[2]Department of Internal Medicine I, Ulm University, Albert-Einstein-Allee 23, 89081 Ulm, Germany
E-Mail: tobias.paust@uni-ulm.de

**Abstract.** Active microrheology is a valuable tool to determine viscoelastic properties of polymer networks. Observing the beads response to the excitation of a reference leads to dynamic and morphological information of the material. In this work we present an expansion of the well-known active two-point microrheology. By measuring the response of multiple particles in the viscoelastic medium in response to the excitation of a reference particle we are able to determine the force propagation in the polymer network.

---

[3] Author to whom any correspondence should be addressed.

## 1. Introduction

The dynamic shear modulus informs about the properties of polymer networks. It can be determined by recording and mathematically transforming the thermal motion of a particle embedded in the viscoelastic medium into the frequency domain. Since no external forces are applied to the particle's motion, this method is named passive microrheology [1], [2], [3], [4]. The resulting shear modulus shows the elastic and diffusive behavior of the investigated medium over the frequency range accessible by the measuring setup. This output is the result of different methods handling the unilateral Laplace transform [5], [6], [7]. By exciting a particle with an oscillation force, the shear modulus at a specific frequency can be determined by the response of the particle. The particle's motion also includes information on the damping and the viscosity of the surrounding medium. This method is known as active microrheology [8], [9], [10]. Both the passive and the active method provide an insight into the storage and the loss modulus of the medium. An extension to the single particle passive method is achieved by the measurement of two or more particles. It was realized that the movement of two particles embedded in a matrix are correlated [11], [12]. The authors showed that the correlation of the movement of two particles in fibronectin leads to calculated viscoelastic parameters in good agreement with classical rheology. Actin networks exhibit a similar correlated movement [13]. It was shown that the single particle and multi-particle technique can lead to equal results [14]. For some networks characteristic differences between one-point and two-point microrheology data were found. It was mentioned by the authors that inhomogeneities could be determined. An overview over these techniques is given in [15] and [16].

## 2. Theoretical aspects

We present a novel active multi-point microrheology method which allows to determine the isotropy of and the force propagation within the viscoelastic medium. Our method does not depend on correlations between particles motions, but analyses the mutual displacements directly. Via the recorded motions of a group of particles located and connected to each other in the viscoelastic medium, it is possible to determine the transfer tensor of motion, the relationship of response amplitudes, phase shifts and frequency changes to higher harmonics. This information depends on the position of the particles and the distances to each other. In a group of particles, one particle - the reference particle $R$ - is excited to sinusoidal oscillations at a specific frequency $\omega$, amplitude $A$ and direction $\theta$. Consequently, the motion of the particle is a superposition of thermal noise and the sinusoidal oscillation. The motion of the response particles is also a superposition of thermal noise, the same sinusoidal function and several terms of the sinusoidal function with doubled, three-fold to $n$-fold frequencies. These are the higher order harmonics whose amplitudes are dependent on the nonlinearity of the system [17] and can be calculated using a Taylor series expansion combined with the addition theorems for sines and cosines. The motion of the response particle can be written as

$$s_n(t) = \frac{a_0}{2} + a_1 \cos(\omega t) + a_2 \cos(2\omega t) + \cdots + a_n \cos(n\omega t) \\ + b_1 \sin(\omega t) + b_2 \sin(2\omega t) + \cdots + b_n \sin(n\omega t) \qquad (1)$$

with $a_0$ to $a_n$ and $b_0$ to $b_n$ being the Fourier coefficients for the individual terms which are dependent on the $k$-fold frequency. The multiplication of the excitation function $f(t)$ with sine and cosine, and its time averaging leads to the Fourier coefficients

$$a_k = \frac{2}{T}\int_0^T f(t)\cos(k\omega t)\,dt$$

(2)

$$b_k = \frac{2}{T}\int_0^T f(t)\sin(k\omega t)\,dt.$$

The response function can be expressed as a sum of sine functions under the assumption that the series converges in time. With the coefficients $a_k$ and $b_k$ the relationship of response amplitudes $x_k$ and phases $\varphi_k$ to the reference can be gained. In all spatial directions the response vector is

$$\vec{r}_{i\theta}(t) = \begin{pmatrix} x_{i0\theta} + x_{i1\theta}\sin(\omega t + \varphi_{xi1\theta}) + \cdots + x_{in\theta}\sin(n\omega t + \varphi_{xin\theta}) \\ y_{i0\theta} + y_{i1\theta}\sin(\omega t + \varphi_{yi1\theta}) + \cdots + y_{in\theta}\sin(n\omega t + \varphi_{yin\theta}) \\ z_{i0\theta} + z_{i1\theta}\sin(\omega t + \varphi_{zi1\theta}) + \cdots + z_{in\theta}\sin(n\omega t + \varphi_{zin\theta}) \end{pmatrix}. \tag{3}$$

In this equation $x_{ik\theta}$, $y_{ik\theta}$ and $z_{ik\theta}$ denote the amplitudes of response particle $i$ for the specific spatial direction $x$, $y$ or $z$ with the excitation direction $\theta$ at harmonic $k$. The phase information for a corresponding $k$ is $\varphi_{xik\theta}$, $\varphi_{yik\theta}$ and $\varphi_{zik\theta}$. The transfer tensor $\overline{A}_i$, which contains information about the material, links both the reference motion and the response motion.

$$\overline{Y}_i = \overline{A}_i \cdot \overline{X}_i$$

with

$$\overline{X}_0 = \begin{pmatrix} x_{011} & x_{012} & \cdots \\ y_{011} & y_{012} & \cdots \\ z_{011} & z_{012} & \cdots \end{pmatrix} \qquad \overline{Y}_i = \begin{pmatrix} x_{i11} & x_{i12} & \cdots \\ y_{i11} & y_{i12} & \cdots \\ z_{i11} & z_{i12} & \cdots \end{pmatrix} \tag{4}$$

with $\overline{X}_0$ the reference matrix and $\overline{Y}_i$ the response matrix of particle $i$. The number of columns of both matrices is dependent on the amount of excited directions. In order to solve the over determined linear equation, the minimum least squares method is taken into account. To gather a unique solution Equation (4) is transposed and left-hand sided multiplied by the reference matrix. If the reference matrix is left-invertible, the matrix product can be inverted and put on the other side of the equation [18]. This leads to

$$\overline{A}_i^T = \left(\overline{X}_0 \overline{X}_0^T\right)^{-1} \cdot \overline{X}_0 \overline{Y}_i^T. \tag{5}$$

The transfer tensor describes the change of the excitation vector to the response vector of the particles.

### 3. Methods and materials

For the determination of the transfer tensor at specific media an external force is needed. This external force was applied with the usage of an optical tweezers. The excitation of the reference particle to oscillations was realized by the sinusoidal motion of the optical trap. An acousto-optical deflector deflected the laser beam to allow the movement of the trap. To ensure beam symmetry, the trapping beam pivots at the back focal plane of the objective. The setup of the optical tweezers and a screenshot of the measurement of several particles in an *in vitro* assembled intermediate filament network is shown in Figure 1.

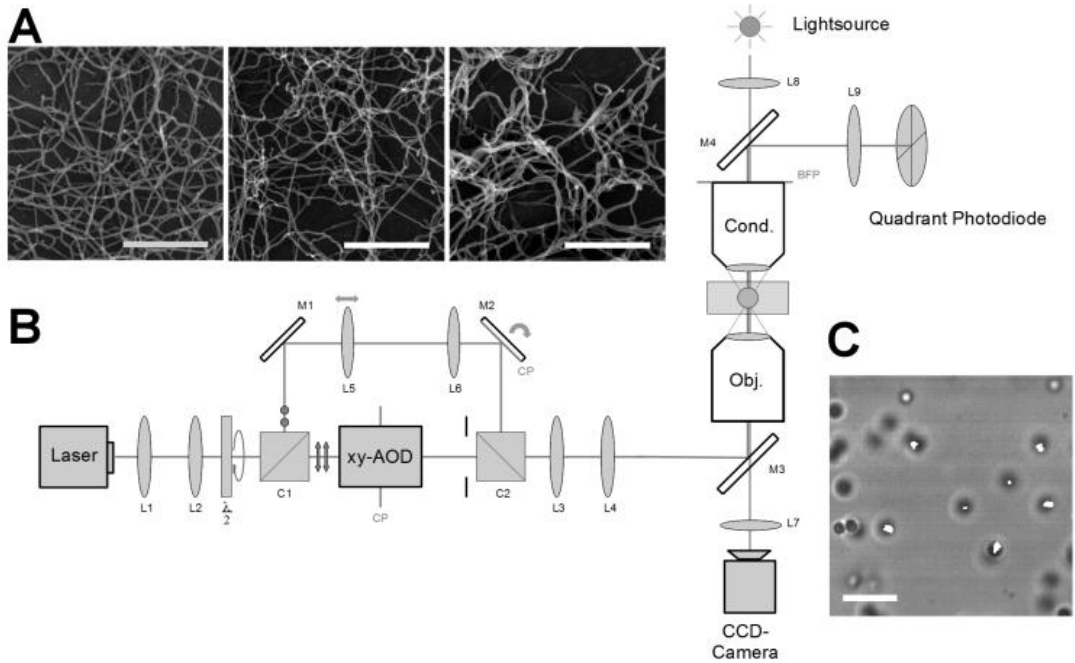

**Figure 1.** A) Three types of *in vitro* assembled keratin networks with different amount of crosslinker (left 0.0 mM, center 0.25 mM, right 2.0 mM $Mg^{2+}$). The scale bar for all three pictures shows a length of 500 nm. B) Setup of the measurement device (optical tweezers). The laser beam is focused into the sample to allow the trapping of micro-spheres. An acousto-optical deflector (AOD) ensures the oscillation of the laser beam with its pivot in the back focal plane of the objective. The CCD high-speed camera records the motion of the micro-spheres embedded in the examined medium. An additional photo-diode is used for the calibration of the trap. C) Image of a microrheology measurement. The white lines show the trajectories of the particle motion. The length of the scale bar is 10 μm.

To validate the described method, two different types of measurements were investigated. The test of the functionality of the setup was performed in bi-distilled water by observing only the trapped particle. The measurements on the viscoelastic medium contained a network of intermediate filaments [19]. For this, human keratins K8 and K18 were synthesized and purified as described by Herrmann et al. [20] and assembled to a network with embedded spherical polystyrene particles [21]. The equipment used for recording the particle motion is described in [21], [7] and [22].

## 4. Simulations

The accuracy of a lock-in-amplifier type detection depends on the noise [23]. Simulations were performed to test how a bad SNR leads to a failure of the method. For this, noise with Gaussian distribution was generated in that way, that the corresponding potential matches to the theoretical potential of the trap with a stiffness $k_{Tr} = 1$ pN/μm. Afterwards, the noise was added to a generated sinusoidal motion with a frequency $f = 10$ Hz and a data length of 16000 points. The calculation of the SNR in the Lock-In method was repeated several times and averaged. Figure 2 shows the dependency of the ratio between expected and calculated amplitude on the excitation amplitude. The inlet depicts the minimal amplitude, when the calculated amplitude is in the range of 1% and 5% compared to the expected one. This minimal amplitude is determined with a 3σ accuracy and dependent on the stiffness of the trap.

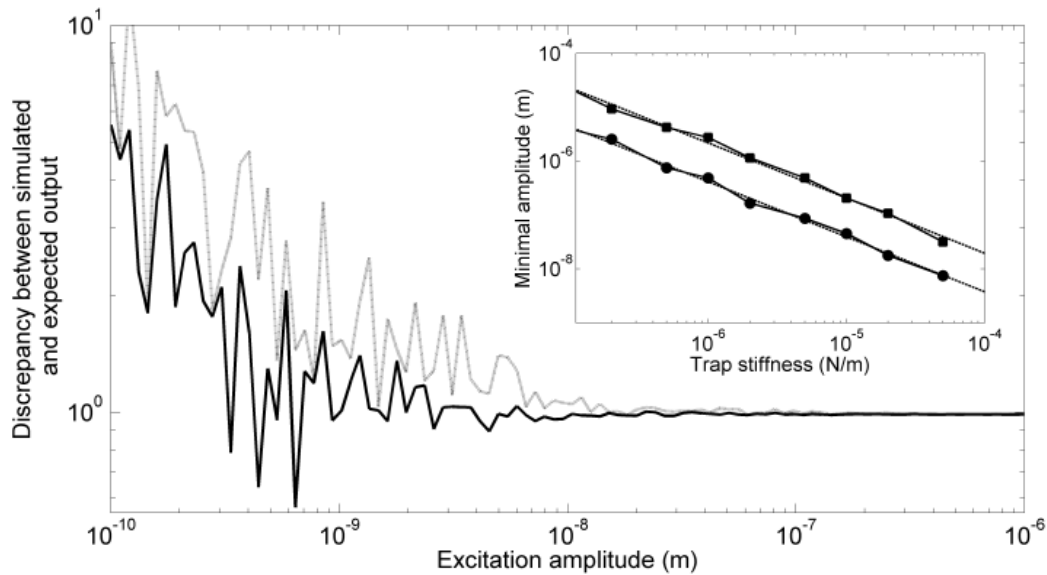

**Figure 2.** Simulations of the accuracy of the Lock-In method. The black line shows the ratio between expected oscillation amplitude and simulated amplitude in presence of noise. In the case when the expected and simulated amplitudes match, the ratio converges to one. Any other case shows a misleading amplitude - the SNR is too low. With the settings of the measurements the ratio increases for oscillation amplitudes smaller than $10^{-8}$ m. The gray line depicts the 5 % deviation with 3σ accuracy. *Inlet:* Minimal amplitude, when the calculated amplitude is in the range of 1% and 5% compared to the expected one (with 3σ accuracy).

At an amplitude of $10^{-8}$ m the calculated amplitude starts to differ from the expected one. For smaller amplitudes this discrepancy increases rapidly. This means that the SNR is too low to gain a meaningful result out of a measurement. In order to improve the SNR and to shift the error to smaller amplitudes, data sets with a length of $10^6$ data points have to be generated. With this change a minimal amplitude of around 1 nm can be achieved. Another possibility to improve the measurement is the change of the trap stiffness. This can be seen in the inlet of Figure 2, where a higher stiffness results in a lower minimal amplitude. To reach the best measurement system, both upper ideas have to be considered. If the number of data points and the trap stiffness is low, the Lock-In method fails because of a too high SNR. This was also tested with the *in vitro* assembled keratin network at a high crosslinker concentration.

## 5. Results

As mentioned above the measurement in bi-distilled water was investigated to confirm the functionality of the setup (data not shown). In that way a sinusoidal motion of the optical trap leads to a similar motion of the particle. A lack of the motion of the particle in the axial direction confirms that the trap focus stays at constant height. Furthermore, the expected response amplitudes for different angles should always be the same as of the trap. This is valid if the oscillation frequency is small so that the viscosity of the medium does not influence the particle motion. Since the response particle phase is always dependent on the reference particle phase, the expected value should lead to zero for bi-distilled water, because the response particle and the reference particle are the same particle. The calculated transfer matrix is predicted to be the identity matrix because of the calculation with equal matrices. The calculation of the amplitudes corresponding to the higher frequency modes leads to zero because a linear response is assumed. The measured particle motion was in accordance with the calculated particle motion via the Lock-In method in the direction of excitation because the Brownian motion was small compared to the oscillations. The diagonal oscillations (45°, 135°) led to amplitudes of $A/\sqrt{2}$. Hence, the measured values were in accordance with theory. The amplitude in non-excited direction always resulted in minor values. This shows that the method works in the case of a viscous fluid.

Afterwards, the transfer matrix was calculated via the reference and response matrix of the observed motions. This led to

$$\overline{X}_0 = \begin{pmatrix} 0.617 & 0.414 & 0.049 & 0.402 \\ 0.030 & 0.422 & 0.632 & 0.348 \\ 0.033 & 0.028 & 0.077 & 0.027 \end{pmatrix} \mu m \qquad \overline{Y}_0 = \begin{pmatrix} 0.617 & 0.414 & 0.049 & 0.402 \\ 0.030 & 0.422 & 0.632 & 0.348 \\ 0.033 & 0.028 & 0.077 & 0.027 \end{pmatrix} \mu m$$

and

$$\overline{A} = \begin{pmatrix} 1.000 & -3.335 \times 10^{-8} & 0.000 \\ 0.000 & 1.000 & 5.189 \times 10^{-8} \\ 0.000 & 3.196 \times 10^{-6} & 1.000 \end{pmatrix}.$$

The columns of matrix $\overline{X}_0$ and $\overline{Y}_0$ contain the amplitudes ($A$=1.275 μm) at an excitation frequency of 10 Hz for the different oscillation angles (0°, 45°, 90° and 135°). The resulting transfer matrix shows the predicted values apart from rounding artifacts which are in the range of $10^{-6}$ to $10^{-8}$.

Similar to the calculations of the transfer matrix in bi-distilled water, the transfer matrices of three types of *in vitro* assembled keratin 8/18 network were determined. The first network was assembled without any additional crosslinker. Network 2 and 3 contained 0.25 mM $Mg^{2+}$ and 1 mM $Mg^{2+}$ as crosslinker. Since an increasing amount of crosslinker evokes a denser and stiffer network [21], the motions of both the reference and response particles decrease. Furthermore, this results in a change of the transfer tensor and the isotropy of the network. Figure 3 shows the motions of reference and response particle in the direction of excitement for the three different networks.

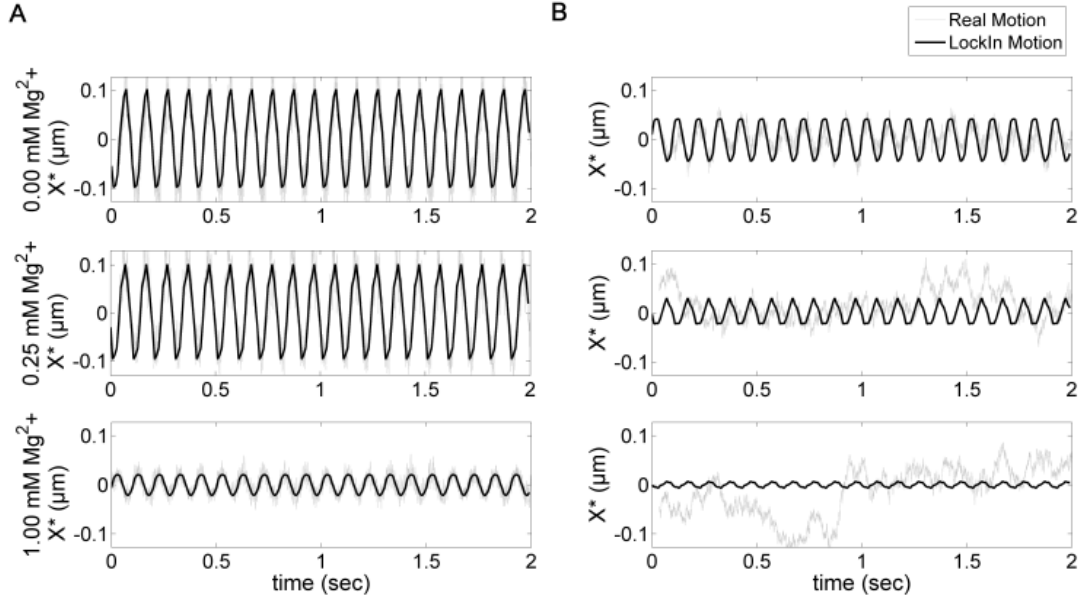

**Figure 3.** Motion of the particles in the network due to the excitation of the oscillating optical trap (A=0.127 µm). The black line denotes the calculated motion via the Lock-In method whereas the gray line is the recorded motion. A) The response motion of the reference particle, for the laser beam oscillating in the same direction with equal strength, dependent on the amount on added crosslinker. The amplitudes decrease due to a denser network. B) The response motion of particles with a distance of 3 µm to the reference particle, dependent on the crosslinker concentration. Since for higher $Mg^{2+}$ concentrations the reference particle shows a small amplitude, the response of a neighboring particle decreases. This implies a decrease of the SNR and introduces errors.

Without any crosslinker the calculated Lock-In motion of the reference particle overlaps with the recorded motion of the CCD camera. A response particle, in a distance of 3 µm to the reference, moves with similar sinusoidal motion as the reference, but with decreased amplitude and phase shift due to the influence of the network. The recorded motion of the camera is in the same range. In the second case, when a crosslinker is added to the network, the amplitude of the reference motion is smaller than the excitation. While at 0.25 mM $Mg^{2+}$ this decrease is negligible, at 1 mM $Mg^{2+}$ only 1/10th of the excitation amplitude deforms the network. At a distance of 3 µm the response motion is reduced because of the decreased excitation and the density of the network. Furthermore, the Brownian motion (recorded motion) becomes larger than the response motion which leads to a decreasing SNR. In the case of 1 mM $Mg^{2+}$ this results in a dramatic increase of calculation errors. The exemplary transfer matrices calculated for particles at a distance of 3 µm for all three cases show that the system is not isotropic.

$$\overline{A}_{0mM} = \begin{pmatrix} 0.442 & 0.206 & 0.025 \\ 0.008 & 0.448 & -0.099 \\ -0.002 & -0.141 & 0.152 \end{pmatrix}, \quad \overline{A}_{0.25mM} = \begin{pmatrix} 0.407 & 0.203 & 0.153 \\ -0.034 & 0.258 & -0.017 \\ 0.343 & -0.161 & 0.377 \end{pmatrix},$$

$$\overline{A}_{1mM} = \begin{pmatrix} 0.207 & -0.089 & 0.027 \\ -0.020 & 0.199 & -0.228 \\ 0.452 & -0.016 & -0.375 \end{pmatrix}$$

In addition to that, the decrease of the diagonal entries for higher crosslinker concentration can be explained by lower response of the particles.

For further calculations the transfer tensors of the networks were separated and averaged dependent on the distance of reference and response particle. Afterwards, a displacement of arbitrary amplitude $a_i$ in the X*- direction, the connecting line between reference and response particle, was chosen to calculate the output vector. The output vector $\vec{r}_O$ was determined via the multiplication of the arbitrary input vector $\vec{r}_I$ and the average transfer tensor $\overline{A}$ at a specific distance of particles.

$$\overline{A}\vec{r}_I = \overline{A}\begin{pmatrix} a_1 \\ 0 \\ 0 \end{pmatrix} = \vec{r}_O \qquad (6)$$

As ratio the absolute values of arbitrary and output vectors were taken into account $|\vec{r}_O| / |\vec{r}_I|$ and weighted by the standard deviation. The error was determined via the standard deviation of the transfer matrices and the propagation of error. The resulting ratios were compared to the theoretically derived values for an elastic isotropic body [24]. Figure 4 depicts the theoretical as well as the measured curves.

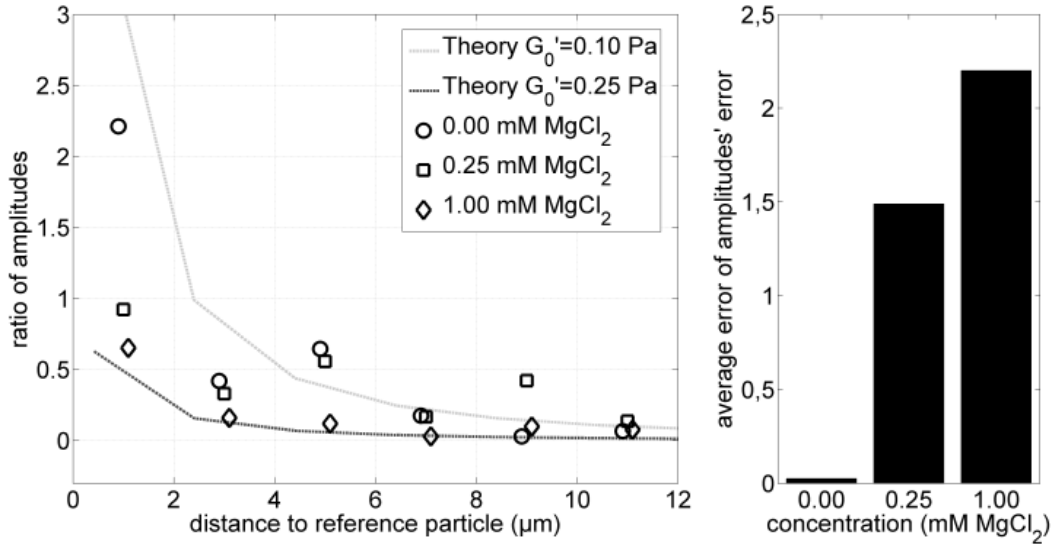

**Figure 4.** *Left:* Ratios of excitation and response dependent on the distance of reference and response particle. Different markers show varying crosslinker concentrations - ○ 0.00 mM MgCl$_2$, □ 0.25 mM MgCl$_2$, ◊ 1.00 mM MgCl$_2$ evoking a denser network for a higher crosslinker concentration [21]. *Right:* Average error of a data point for different crosslinker concentrations. Due to a low SNR the uncertainty of the data increases which leads to meaningless results (also shown in Figure 2.).

For larger distances to the reference particle, the ratio of amplitudes decreased which means that the displacement of the particle got smaller with higher distance between reference and response particle (Figure. 4A). This behavior is comparable to the theoretical predictions. For the measurements at 0 mM Mg$^{2+}$ this can be seen for both directions - for excitation in X* and Y*, perpendicular direction to the connecting line (data not shown), - but it is difficult to recognize a faster decrease of the amplitude ratio with larger distance for the lateral excitation. The larger the distance, the higher the error became because the deformation propagation was highly dependent on the morphology of the network. Hence, for larger distances the probability that the filaments channel the force not directly to the response particle was higher, especially if the network mesh sizes were large. The theoretical curve is dependent on the chosen shear modulus. In these calculations the shear modulus determined via passive microrheology [21], [7] was taken into account. With a plateau modulus of $G_0' = 0.02$ Pa, the theoretical curve was higher than the curve of the measurements. If a theoretical value of $G_0' = 0.1$ Pa was applied to the theoretical calculations, the curve matched better to the measured data, which indicates that, for the moment, both methods - the passive one and the active one - are not comparable.

For the measurements with 1.0 mM $Mg^{2+}$ the prediction is that the response of the network is lower for short distances, assuming that the deformation force is the same as above. The displacement of positions at larger distances is therefore hardly appreciable due to a low deformation and a decreasing SNR. The low SNR introduces errors and makes the results meaningless. A network containing 0.25 mM $Mg^{2+}$ crosslinker shows ratios of amplitudes which are located between the other measurements. That means on the one hand that the network can be deformed easier than at 1mM $Mg^{2+}$, and on the other hand that the SNR is higher so that the errors became less effective. For all measurements the errors are shown in Figure 4B.

It is obvious that the errors were very high for the 1 mM $Mg^{2+}$ network. This was on the one hand due to the low response amplitudes of the reference particle which deformed the network. That led to a high ratio of reference SNR and response SNR which introduced high entries at the transfer tensor. From measurement to measurement the entries in the transfer matrices varied a lot and this resulted in the high uncertainties. On the other hand it was observed in [21] that the higher the salt concentration is, the more often bundles occur in the sample. Then the assumption of an isotropic medium is not valid and for every position it is not known if the response particle is located in a bundle or beside the bundle. Thus, the comparison of the specific directions of excitement and the fit to a theoretical curve are not reasonable for this case. To improve the results and decrease the error it might be recommended to additionally investigate the amplitudes of the higher harmonics of the measurement. If those amplitudes are comparable to the first harmonic, nonlinear effects play a role, which complicate the interpretations.

With the simulation we can calculate the influence of the SNR to the ratio of amplitudes and estimate a lower limit of SNR, when the error leads to meaningless results. In the case of 1.0 mM $Mg^{2+}$ this limit was exceeded so that the response of the particles was not recorded, but the Brownian motion and surrounding noise.

## 6. Conclusion

With the novel active method new insights into the dynamics of cytoskeletal networks can be gained. Properties like the response amplitude propagation through the network, the isotropy of the network or the phase change while deforming the network can be determined by one set of measurements. These parameters help in the understanding of network architecture and can be used to estimate the force propagation on external stress. In living cells, this force propagation plays an important role for the cells motion, reaction to external influences and transport of vesicles. In this work we showed the theoretical idea behind the method, the experimental implementation and in addition its limits via calculations of the SNR. These preliminary results open a wide field of applications in all kinds of different viscoelastic media.

**Acknowledgements**

This project was supported by the German Research Association (DFG) (SFB 569, MA 1297/10-1, BE 2339/3-1) and the Landesstiftung Baden-Württemberg. We thank Harald Herrmann and his group from the DKFZ Heidelberg, Germany for providing the keratin monomers. The help of Ulla Nolte for the preparation and the discussions with Kay Gottschalk are gratefully acknowledged.